\documentclass[a4paper,10pt]{article}

\title{Lev Landau and the problem of singularities in cosmology}
\author{I M Khalatnikov$^{1}$ and A Yu Kamenshchik$^{1,2}$}
\date{}
\begin{document}

\maketitle
\hspace{-5mm}$^1$L.D. Landau Institute for Theoretical Physics of Russian Academy of Sciences,
Kosygin str. 2, 119334 Moscow, Russia\\
$^2$Dipartimento di Fisica and Istituto Nazionale di Fisica Nucleare, via Irnerio 46, 40126 Bologna, Italy
\begin{abstract}
We consider different aspects of the problem of cosmological singularity such as the BKL oscillatory approach
to singularity, the new features of the cosmological dynamics in the neighbourhood of the singularity
in multidimensional and superstring cosmological models and their connections with such a modern branch of
mathematics as infinite-dimensional Lie algebras. Besides, we consider some new types of cosmological singularities
which were widely discussed during last decade after the discovery of the phenomenon of cosmic acceleration.
\end{abstract}

\section{Introduction}
Many years ago, in conversations with his students, Lev Davidovich Landau
used to say that there were three problems most important for the
theoretical physics:
the problem of cosmological singularity, the problem of phase transitions
and the problem of superconductivity \cite{Landau}.
Now we know that the great breakthrough was achieved in the explanation of
phenomena of superconductivity \cite{superconduct} and phase transitions \cite{phase}.
The problem of cosmological singularity was widely studied during the last 50 years
and many important results were obtained, but it still conserves some intriguing
aspects. Moreover some quite unexpecting facets of the problem of cosmological
singularity were discovered.

In our review, published 10 years ago \cite{ufn} in the issue of this journal,
dedicated to the 90th anniversary of L D  Landau birth, we had discussed some
questions connected with the problem of singularity in cosmology. In the present
paper we shall dwell on relations between well-known old results of these studies
and the new developments of this area.

To begin with, let us remember that R Penrose and S Hawking \cite{Pen-Hawk} proved
the impossibility of indefinite continuation of geodesics under certain conditions.
This was interpreted as pointing to the existence of a sigularity in the general
solution of the Einstein equations. These theorems, however, did not allow for finding
the particular analytical structure of the singularity.
The analytical behaviour of the general solutions of the Einstein equations in the
neighborhood of singularity was investigated in the papers by E M Lifshitz and I M Khalatnikov
\cite{LK0,LK,LK1} and V A Belinsky, E M Lifshitz and I M Khalatnikov \cite{BKL,BKL1,BKL2}.
These papers revealed the enigmatic  phenomenon of an oscillatory approach to the singularity
which has become known as the {\it Mixmaster Universe} \cite{Misner}.
The model of the closed homegeneous but anysotropic universe with three degrees of freedom
(Bianchi IX cosmological model) was used to demonstrate that the universe approaches the singularity
in such a way that its contraction along two axes is accompanied by an expansion with respect to
the third axis, and the axes change their roles according to a rather complicated law which reveals
chaotical behavior \cite{BKL1,BKL2,LLK,five}.

The study of the dynamics of the universe in the vicinity of the cosmological singularity
has become a explodingly developing field of modern theoretical and mathematical physics.
First of all we would like to mention the generalization of the study of the oscillatory approach
to the cosmological singularity in multidimensional cosmological models. It was noticed \cite{multi}
that the  approach to the cosmological singularity in the multidimensional (Kaluza-Klein type)
cosmological models has a chaotic character in the spacetimes whose dimensionality is not higher then
ten, while in the spacetimes of higher dimensionalities a universe after undergoing a finite number
of oscillations enters into monotonic Kasner-type contracting regime.

The development of cosmological studies based on  superstring models has reavealed some new aspects
of the dynamics in the vicinity of the singularity \cite{DHN}.
First, it was shown that in these models exist mechanisms of changing of Kasner epochs provoked not by
the gravitational interactions but by the influence of other fields present in these theories.
Second, it was proved that the cosmological models based on six main superstring models plus $D=11$ supergravity
 model exhibit the chaotical oscillatory approach towards the singularity.
Third, the connection between cosmological models, manifesting the oscillatory approach towards singularity and
a special subclass of infinite-dimensional Lie algebras \cite{Kac} - the so called hyperbolic Kac-Moody algebras was
discovered.

Another confirmation of the importance of the problem of singularity in general relativity has
come from the observational cosmology. At the end of 90th the study of the relation between the luminosity and redshift
of supernovae of the type Ia has reavealed that the modern Universe is expanding with an acceleration \cite{cosmic}.
To provide such an acceleration it is necessary to have a particular substance which was dubbed ``dark energy'' \cite{dark}.
The main feature of this kind of matter is that it should have a negative pressure $p$ such that $\rho + 3p < 0$,
where $\rho$ is the energy density. The simplest kind of this matter is the cosmological constant for which
$p = -\rho$. The so called standard or $\Lambda$CDM cosmological model is based on the cosmological constant, whose
energy density is responsable for roughly 70 percents of the general energy density of the Universe, while the rest
is occupied by a dust-like matter both baryonic (approximately 4 percents) and dark one. This model is in a good agreement with
the observations, however other candidates for the role of dark energy are intensively studied and new observations can give
some surprises alredy in the nearest years. First of all let us notice that some observations \cite{obs-phan} point out on the possible existence
of the so called superacceleration, which is connected with the presence of the phantom  dark energy \cite{phantom},
characterized by the inequality $p < -\varepsilon$. The universe, filled with such kind of dark energy, at some conditions can encounter a very
particular cosmological singularity - the Big Rip \cite{Rip}.
When an expanding universe encounters this singularity it has an infinite cosmological
radius and an infinite value of the Hubble variable. Earlier the possibility of this type of singularity was discussed in paper \cite{Star-Rip}.

Study of different possible candidates for the role of dark energy
has stimulated the elaboration of general theory of possible
cosmological singularities \cite{Brake,sudden,finite,boost,brane}.
It is remarkable that while the ``traditional'' Big Bang or Big
Crunch singularities are associated with vanishing size of a
universe, i.e. with a universe squeezed to a point, these new
singularities occur at finite or infinite value of the cosmological
radius. The physical processes taking place in the vicinity of such
singularities can have rather exotic features and their study is of
great interest. Thus, we see that the development of both the
theoretical and the observational branches of cosmology has
confirmed the importance of the problem of singularity in general
relativity mentioned by Landau many years ago.

The structure of the paper is the following:
in Sec. 2 we briefly discuss the  Landau theorem about the singularity
which was not published in a separate paper and was reported in the book \cite{LL} and
in the review \cite{LK0};
in Sec. 3 we shall recall the main features of the oscillatory approach to the singularity in relativistic cosmology;
Sec. 4 will be devoted to the modern development of the BKL ideas and methods, including the dynamics in the presence of a massless scalar field, multidimensional cosmology,
superstring cosmology and the correspondence between chaotic cosmological dynamics and hyperbolic Kac-Moody algebras;
in Sec. 5 we describe some new types of cosmological singularities and in Sec. 6 we shall present some concluding remarks.

\section{Landau theorem about the singularity}
Let us consider the synchronous reference frame with the metric
\begin{equation}
ds^2 = dt^2 - \gamma_{\alpha\beta}dx^{\alpha}dx^{\beta},
\label{synchron}
\end{equation}
where $\gamma_{\alpha\beta}$ is the spatial metric. L D  Landau pointed out
that the determinant $g$ of the metric tensor in a synchronous reference system
must go to zero at some finite time provided some simple conditions on the equation
of state are satisfied. To prove this statement it is convenient to write down the $0-0$
component of the Ricci tensor as
\begin{equation}
R_0^0  = -\frac12 \frac{\partial K_{\alpha}^{\alpha}}{\partial t}  -
\frac14 K_{\alpha}^{\beta} K_{\beta}^{\alpha},
\label{R00}
\end{equation}
where $K_{\alpha\beta}$ is the extrinsic curvature tensor  defined as
\begin{equation}
K_{\alpha\beta} = \frac{\partial \gamma_{\alpha\beta}}{\partial t}
\label{extrinsic}
\end{equation}
and the spatial indices are raised and lowered by the spatial metric
$\gamma_{\alpha\beta}$. The Einstein equation for $R_0^0$ is
\begin{equation}
R_0^0 = T_0^0 - \frac12 T,
\label{Einstein}
\end{equation}
where the energy-momentum tensor is
\begin{equation}
T_{i}^{j} = (\rho + p)u_i u^j - \delta_i^j p,
\label{tensorem}
\end{equation}
where $\rho, p$ and $u_i$ are the energy density, the pressure and the
four-velocity respectively. The quantity $T_0^0 - \frac12 T$ in the right-hand side
of equation (\ref{Einstein}) is
\begin{equation}
T_0^0 - \frac12 T = \frac12 (\rho + 3p) + (\rho + p) u_{\alpha}u^{\alpha},
\label{tensorem1}
\end{equation}
and is positive provided
\begin{equation}
\rho + 3p > 0.
\label{energodom}
\end{equation}
Thus, from Eq. (\ref{Einstein}) follows that
\begin{equation}
\frac12 \frac{\partial K_{\alpha}^{\alpha}}{\partial t}  +
\frac14 K_{\alpha}^{\beta} K_{\beta}^{\alpha} \leq 0.
\label{ineq}
\end{equation}
Because of the algebraic inequality
\begin{equation}
K_{\alpha}^{\beta} K_{\beta}^{\alpha}  \geq \frac13
(K_{\alpha}^{\alpha})^2
\label{ineq1}
\end{equation}
we have
\begin{equation}
\frac{\partial K_{\alpha}^{\alpha}}{\partial t}  +  \frac16
(K_{\alpha}^{\alpha})^2 \leq 0
\label{ineq2}
\end{equation}
or
\begin{equation}
\frac{\partial}{\partial t} \frac{1}{K_{\alpha}^{\alpha}} \geq \frac16.
\label{ineq3}
\end{equation}
If $K_{\alpha}^{\alpha} > 0$ at some moment of time, then if $t$ decreases,
the quantity $\frac{1}{K_{\alpha}^{\alpha}}$ decreases to zero within a finite time.
Hence $K_{\alpha}^{\alpha}$ goes to $+\infty$ and that means because of the identity
\begin{equation}
K_{\alpha}^{\alpha} = \gamma^{\alpha\beta}\frac{\partial g_{\alpha\beta}}{\partial t}
= \frac{\partial}{\partial t} \ln g
\label{det}
\end{equation}
that the determinant $g$ goes to zero (no faster than $t^6$ according to the inequality
(\ref{ineq3})).  If at the initial moment $K_{\alpha}^{\alpha} < $, then the same result
will be obtained for the increasing time. The similar result
was obtained by Raychaudhuri \cite{Ray} for the case of dust-like matter and by Komar
\cite{Komar}.

This result does not prove that it is inevitable that there exists a true physical
singularity, which is one belonging to the spacetime itself and is not connected
with the character of the chosen reference system. However, it played an important role
stimulating the discussion about the existence and generality of the
singularities in cosmology.
Notice that the condition of the energodominance (\ref{energodom}) used for the
proof of the Landau theorem appears also in the proof of the Penrose and Hawking
singularity theorem \cite{Pen-Hawk}. Moreover, the breakdown of this condition is
necessary for an explanation of the phenomenon of cosmic acceleration.

The Landau theorem is deeply connected with the appearance of
caustics studied by E M Lifshitz I M Khalatnikov and V V Sudakov
\cite{LSK} and discussed between them and L D Landau in 1961. Trying
to construct geometrically the synchronous reference frame, one
start from the three-dimensional Cauchy surface and design the
family of geodesics orthogonal to this this surface. The length
along these geodesics serves as the time measure. It is known that
these geodesics intersect on some two-dimensional caustic surface.
This geometry constructed for the empty space is valid also in the
presence of dust-like matter ($p = 0$). Such matter, moving along
the geodesics concentrates on caustics, but the growth of density
cannot be unbounded because the arising pressure destroys the
caustics \footnote{In an empty space the caustic is a mathematical,
but not a physical singularity. This follows simply from the fact
that we can always shift its location by changing the initial Cauchy
surface.}. This question was studied by L P Grishchuk \cite{Grish}.
Later V I Arnold, S F Shandarin and Ya B Zeldovich \cite{ASZ} have
used the caustics for the explanation of the initial clustering of
the dust, which though does not create physical singularities, is
nevertheless responsible for the creation of the so called pancakes.
These pancakes represent the initial stage of the development of the
large scale structure of the universe.

\section{Oscillatory approach to the singularity in relativistic
cosmology}
One of the first exact solutions found in the framework of general
relativity was the Kasner solution \cite{Kasner} for the Bianchi-I
cosmological model representing gravitational field in an empty space
with Euclidean metric depending on time according to the formula
\begin{equation}
ds^{2} = dt^{2} - t^{2p_{1}}dx^{2} - t^{2p_{2}}dy^{2}
-t^{2p_{3}}dz^{2},
\label{Kasner}
\end{equation}
where the exponents $p_{1}, p_{2}$ and $p_{3}$ satisfy the relations
\begin{equation}
p_{1}+ p_{2}+ p_{3} = p_{1}^{2}+ p_{2}^{2}+ p_{3}^{2} = 1.
\label{exponents}
\end{equation}
Choosing the ordering of exponents as
\begin{equation}
p_{1} < p_{2} < p_{3}
\label{ordering}
\end{equation}
one can parameterize them as \cite{LK0}
\begin{equation}
p_{1} = \frac{-u}{1+u+u^{2}},\;
p_{2} = \frac{1+u}{1+u+u^{2}},\;
p_{3} = \frac{u(1+u)}{1+u+u^{2}}.
\label{u-define}
\end{equation}
As the parameter $u$ varies in the range $u \geq 1$,
$p_{1}, p_{2}$ and $p_{3}$ assume all their permissible values
\begin{equation}
-\frac{1}{3} \leq p_{1} \leq 0,\; 0 \leq p_{2} \leq \frac{2}{3},\;
\frac{2}{3} \leq p_{3} \leq 1.
\label{range0}
\end{equation}
The values $u < 1$ lead to the same range of values of
$p_{1},p_{2},p_{3}$ since
\begin{equation}
p_{1}\left(\frac{1}{u}\right) = p_{1}(u),\;
p_{2}\left(\frac{1}{u}\right) = p_{3}(u),\;
p_{3}\left(\frac{1}{u}\right) = p_{2}(u).
\label{u-define1}
\end{equation}
The  parameter $u$ introduced in early sixties has
appeared very useful and its properties attract attention of researchers
in different contests. For example, in the recent paper \cite{LK-Pet} a connection was
established between the Lifshitz-Khalatnikov parameter $u$ and the invariants, arising
in the context of the Petrov's classification of the Einstein spaces \cite{Petrov}.

In the case of Bianchi-VIII or Bianchi-IX cosmological models
the Kasner regime (\ref{Kasner}),(\ref{exponents}) ceased to be an
exact solution of Einstein equations, however one can design the
generalized Kasner solutions \cite{LK}-\cite{BKL2}.
It is possible to construct some kind of perturbation theory where
exact Kasner solution (\ref{Kasner}),(\ref{exponents}) plays role of
zero-order approximation while role of perturbations play those terms
in Einstein equations which depend on spatial curvature tensors
(apparently, such terms are absent in Bianchi-I cosmology). This
theory of perturbations is effective in the vicinity of singularity
or, in other terms, at $t \rightarrow 0$. The remarkable feature of
this perturbations consists in the fact that they imply the
transition from the Kasner regime with one set of parameters
to the Kasner regime with another one.

The metric of the generalized Kasner solution in a synchronous
reference system can be written in the form
\begin{equation}
ds^{2} = dt^{2} - (a^{2}l_{\alpha}l_{\beta}
+b^{2}m_{\alpha}m_{\beta}+c^{2}n_{\alpha}n_{\beta})dx^{\alpha}
dx^{\beta},
\label{metric}
\end{equation}
where
\begin{equation}
a = t^{p_{l}},\;b = t^{p_{m}},\;c = t^{p_{n}}.
\label{exponents1}
\end{equation}
The three-dimensional vectors $\vec{l}, \vec{m}, \vec{n}$ define the
directions along which the spatial distances vary with time according
to the power laws (\ref{exponents1}).
Let $p_{l} = p_{1}, p_{m} = p_{2}, p_{n} = p_{3}$ so that
\begin{equation}
a \sim t^{p_{1}},\;b \sim t^{p_{2}},\;c \sim t^{p_{3}},
\label{exponents2}
\end{equation}
i.e. the Universe is  contracting in directions given by vectors
$\vec{m}$ and $\vec{n}$ and is expanding along $\vec{l}$.
It was shown \cite{BKL1} that the perturbations caused by
spatial curvature terms make the variables $a, b$ and $c$ to undergo
transition to another Kasner regime characterized by the following
formulae:
\begin{equation}
a \sim t^{p_{l}'},\;b \sim t^{p_{2}'},\;c \sim t^{p_{3}'},
\label{Kasner1}
\end{equation}
where
\begin{equation}
p_{l}' = \frac{|p_{1}|}{1-2|p_{1}|},\;
p_{m}' = -\frac{2|p_{1}|-p_{2}}{1-2|p_{1}|},\;
p_{n}' = -\frac{p_{3}-2|p_{1}|}{1-2|p_{1}|}.
\label{Kasner2}
\end{equation}

Thus, the effect of the perturbation is to replace one ``Kasner
epoch'' by another so that the negative power of $t$ is transformed
from the $\vec{l}$ to the $\vec{m}$ direction. During the transition
the function $a(t)$ reaches a maximum and $b(t)$ a minimum. Hence,
the previously decreasing quantity $b$ now increases, the quantity
$a$ decreases and $c(t)$ remains a decreasing function. The
previously increasing perturbation caused the transition from regime
(\ref{exponents2}) to that (\ref{Kasner1}) is damped and eventually
vanishes. Then other perturbation begins grow which leads to a new
replacement of one Kasner epoch by another, etc.

We would like to emphasize  that namely the fact that perturbation
implies such a change of dynamics which extinguishes it gives us an
opportunity to use perturbation theory so successfully. Let us note
that the effect of changing of the Kasner regime exists already in
the cosmological models more simple than those of Bianchi IX and
Bianchi VIII. As a matter of fact in the Bianchi II universe there
exists only one type of perturbations, connected with the spatial
curvature and this perturbation makes one change of Kasner regime
(one bounce). This fact was known to E M Lifshitz and I M
Khalatnikov at the beginning of sixties and they have discussed this
topic with L D Landau (just before the tragic accident) who has
appreciated it highly. The  results describing  the dynamics of the
Bianchi IX model were reported by I M Khalatnikov in his talk given
in January 1968 in Henri Poincare Seminar, in Paris. John A Wheeler
who was present there pointed out that the dynamics of the Bianchi
IX universe represents a non-trivial example of the chaotic
dynamical system. Later Kip Thorn has distributed the preprint with
the text of this talk.

Coming back to the rules governing  the bouncing of the negative
power of time from one direction to another one can show that they
could be conveniently expressed by means of the parameterization
(\ref{u-define}):
\begin{equation}
p_{l} = p_{1}(u),\; p_{m} = p_{2}(u),\; p_{n} = p_{3}(u)
\label{u-define2}
\end{equation}
and then
\begin{equation}
p_{l}' = p_{2}(u-1),\; p_{m}' = p_{1}(u-1), \;p_{n}' = p_{3}(u-1).
\label{u-define3}
\end{equation}
The greater of the two positive powers remains positive.

The successive changes (\ref{u-define3}), accompanied by a bouncing
of the negative power between the directions $\vec{l}$ and $\vec{m}$,
continue as long as the integral part of $u$ is not exhausted,
i.e. until $u$ becomes less that one. Then, according to  Eq. (\ref
{u-define1}) the value $u < 1$ transforms into $u > 1$, at this
moment either the exponent $p_{l}$ or $p_{m}$ is negative and $p_{n}$
becomes smaller one of the two positive numbers ($p_{n} = p_{2}$).
The next sequence of changes will bounce the negative power between
the directions $\vec{n}$ and $\vec{l}$ or $\vec{n}$ and $\vec{m}$.
Let us emphasize that the usefulness of the Landau-Khalatnikov
parameter $u$ is connected with the circumstance that it allows
to encode rather complicated laws of transitions between different
Kasner regimes (\ref{Kasner2}) in such simple rules as
$u \rightarrow  u - 1$ and
 $u \rightarrow  \frac{1}{u}$.

Consequently, the evolution of our model towards a singular point
consists of successive periods (called eras) in which distances along
two axes oscillate and along the third axis decrease monotonically,
the volume decreases according to a law which is near to $\sim t$. In
the transition  from one era to another, the axes along which the
distances decrease monotonically are interchanged. The order in which
the pairs of axes are interchanged and the order in which eras of
different lengths follow each other acquire a stochastic character.

To every ($s$th) era corresponds a decreasing sequence of values of
the parameter $u$.  This sequence has the form $u_{max}^{(s)},
u_{max}^{(s)}-1,\ldots,u_{min}^{(s)}$, where $u_{min}^{(s)} < 1$.
Let us introduce the following notation:
\begin{equation}
u_{min}^{(s)} = x^{(s)},\; u_{max}^{(s)} = k^{(s)} + x^{(s)}
\label{era}
\end{equation}
i.e. $k^{(s)} = [u_{max}^{(s)}]$ (the square brackets denote
the greatest integer $\leq u_{max}^{(s)}$). The number $k^{(s)}$
defines the era length. For the next era we obtain
\begin{equation}
u_{max}^{(s+1)} = \frac{1}{x^{(s)}},\;
k^{(s+1)} = \left[\frac{1}{x^{(s)}}\right].
\label{era1}
\end{equation}

The ordering with respect to the length of $k^{(s)}$ of the
successive eras (measured by the number of Kasner epochs contained in
them) acquires asymptotically a stochastic character . The random
nature of this process arises because of the rules
(\ref{era})--(\ref{era1}) which define the transitions from one era
to another in the infinite sequence of values of $u$. If all this
infinite sequence begins since some initial value $u_{max}^{(0)}
= k^{(0)} + x^{(0)}$, then the lengths of series $k^{(0)},
k^{(1)},\ldots$ are numbers included into an expansion of a
continuous fraction:
\begin{equation}
k^{(0)} + x^{(0)} = k^{(0)} + \frac{1}{k^{(1)} +
\frac{1}{k^{(2)}+\cdots}} .
\label{fraction}
\end{equation}

We can describe statistically this sequence
of eras if we consider instead of a given initial value
$u_{max}^{(0)} = k^{(0)} + x^{(0)}$ a distribution of $x^{0)}$ over
the interval $(0,1)$ governed by some probability law. Then we also
obtain some distributions of the values of $x^{(s)}$ which terminate
every $s$th series of numbers. It can be shown that with increasing
$s$, these distributions tend to a stationary (independent of $s$)
probability distribution $w(x)$ in which the initial value $x^{(s)}$
is completely ``forgotten'':
\begin{equation}
w(x) = \frac{1}{(1+x) \ln 2}.
\label{distrib}
\end{equation}
It follows from Eq. (\ref{distrib}) that the probability distribution
of the lengths of series $k$ is given by
\begin{equation}
W(k) = \frac{1}{\ln 2} \ln \frac{(k+1)^{2}}{k(k+2)}.
\label{distrib1}
\end{equation}

Moreover, one can calculate in an exact manner probability
distributions for other parameters describing successive
eras such as parameter $\delta$ giving relation between the
amplitudes of logarithms of functions $a, b, c$ and logarithmic time
\cite{five}.

Thus, we have seen from the results of statistical analysis of
evolution in the neighbourhood of singularity \cite{LLK} that
the stochasticity and probability distributions of parameters arise
already in classical general relativity.

At the end of this section a historical remark is in order. The
continuous fraction (\ref{fraction}) was shown in 1968 to I M
Lifshitz (L D Landau has already left) and he immediately noticed
that one can derive the formula for a stationary distribution of the
value of $x$ (\ref{distrib}). Later it becomes known that this
formula was derived in nineteenth century by Karl F  Gauss, who had
not published it but had described it in a letter to one of
colleagues.

\section{Oscillatory approach to the singularity: modern development}
The oscillatory approach to the cosmological singularity
described in the preceding section was developed  for the case of an empty
spacetime. It is not difficult to understand that if one considers the universe
filled with a perfect fluid with the equation of state $p = w\rho$, where $p$ is the pressure,
$\rho$ is the energy density and $w < 1$, then the presence of this matter
cannot change the
dynamics in the vicinity of the singularity. Indeed, using the energy conservation equation
one can show that
\begin{equation}
\rho = \frac{\rho_0}{(abc)^{w+1}} = \frac{\rho_0}{t^{w+1}},
\label{matter}
\end{equation}
where $\rho_0$ is a positive constant. Thus, the term representing the matter in the Einstein
equations behaves like $\sim 1/t^{1+w}$ and at $t \rightarrow 0$ is weaker than the terms
of geometrical origin coming from the time derivatives of the metric, which behaves like $1/t^2$,
let alone the perturbations due to the presence of spatial curvature, responsable for changes of
a Kasner regime,which behave like $1/t^{2 + 4|p_1|}$. However, the situation changes drastically,
if the parameter $w$ is equal to one, i.e. the pressure is equal to the energy density. Such kind
of matter is called ``stiff matter'' and can be represented by a massless scalar field.
In this case $\rho \sim 1/t^2$ and the contribution of matter is of the same order as main
terms of geometrical origin.  Hence, it is necessary to find a Kasner type solution, taking
into account the presence of terms, connected with the presence of the stiff matter
(a massless scalar field). Such study was carried out in paper \cite{Bel-Khal}.
It was shown that again the scale factors $a,b$ and $c$ can be represented as $t^{2p_1},
t^{2p_2}$ and $t^{2p_3}$ respectively, where the Kasner indices satisfy the relations:
\begin{equation}
p_1 + p_2 + p_3 = 1,\ \ p_1^2 + p_2^2 + p_3^2 = 1 - q^2,
\label{matter1}
\end{equation}
where the number $q^2$ reflects the presence of the stiff matter and is  bounded by
\begin{equation}
q^2 \leq \frac23.
\label{matter2}
\end{equation}
One can see that if $q^2 > 0$, then exist combinations of the positive Kasner indices,
satisfying the relations (\ref{matter1}).
Moreover, if $q^2 \geq \frac12$ only sets of three positive Kasner indices can satisfy
the relations (\ref{matter1}).
If a universe finds itself in a Kasner regime with
three positive indices, the perturbative terms, existing due to the spatial curvatures
are too week to change this Kasner regime, and thus, it becomes stable. That means,
that in the presence of the stiff matter, the universe after a finite number of changes of
Kasner regimes finds itself in a stable regime and oscillations
stop. Thus, the massless scalar field plays ``anti-chaotizing'' role in the process of the
cosmological evolution \cite{Bel-Khal}.
One can use the Lifshitz-Khalatnikov parameter also in this case. The Kasner indices
satisfying the relations (\ref{matter1}) are conveniently represented as \cite{Bel-Khal}
\begin{eqnarray}
&&p_1 = \frac{-u}{1+u+u^2}, \nonumber \\
&&p_2 = \frac{1+u}{1+u+u^2}\left[u-\frac{u-1}{2}(1-(1-\beta^2)^{1/2})\right], \nonumber \\
&&p_3 = \frac{1+u}{1+u+u^2}\left[1+\frac{u-1}{2}(1-(1-\beta^2)^{1/2})\right], \nonumber \\
&&\beta^2 = \frac{2(1+u+u^2)^2}{(u^2-1)^2}.
\label{u-new}
\end{eqnarray}
The range of the parameter $u$ now is $-1 \leq u \leq 1$, while the admissible values
of the parameter $q$ at some given $u$ are
\begin{equation}
q^2 \leq \frac{(u^2-1)^2}{2(1+u+u^2)^2}.
\label{range}
\end{equation}
One can easily show that after one bounce the value of the parameter
$q^2$ changes according to the rule
\begin{equation}
q^2 \rightarrow q'^2  = q^2 \times \frac{1}{(1+2p_1)^2} > q^2.
\label{q}
\end{equation}
Thus, the value of the parameter $q^2$ grows and, hence, the probablity
to find all the three Kasner  indices to be positive  increases. It confirms again the statement
that after a finite number of bounces the universe in the presence of
the massless scalar field finds itself in the Kasner regime with three positive
indices and the oscillations stop.

In the second half of eightees a series of papers was published \cite{multi}, where
were studied the solutions of Einstein equations in the vicinity of singularity for
$d+1$-dimensional spacetimes. The multidimensional analog of a Bianchi-I universe
was considered, where the metric is a generalized Kasner metric:
\begin{equation}
ds^2 = dt^2 - \sum_{i=1}^{d} t^{2p_i}dx^{i2},
\label{gen-Kas}
\end{equation}
where the Kasner indices $p_i$ satisfy the conditions
\begin{equation}
\sum_{i=1}^{d} p_{i} = \sum_{i=1}^{d}p_{i}^2 = 1.
\label{gen-Kas1}
\end{equation}
In the presence of spatial curvature terms the transition from one Kasner epoch to another occurs
and this transition is described by the following rule: the new Kasner exponents are equal to
\begin{equation}
p_1', p_2',\ldots,p_d' = {\rm ordering\  of}\  (q_1,q_2,\ldots,q_d),
\label{new-Kas}
\end{equation}
with
\begin{eqnarray}
&&q_1 = \frac{-p_1 - P}{1+2p_1+P},\ q_2 = \frac{p_2}{1+2p_1+P},\ldots,\nonumber \\
&&q_{d-2} = \frac{p_{d-2}}{1+2p_1+P},\ q_{d-1} = \frac{2p_1+P+p_{d-1}}{1+2p_1+P},\nonumber \\
&&q_d = \frac{2p_1+P+p_d}{1+2p_1+P},
\label{new-Kas1}
\end{eqnarray}
where
\begin{equation}
P = \sum_{i=2}^{d-2} p_i.
\end{equation}
However, such a transition from one Kasner epoch to another occurs if at least one
of the numbers $\alpha_{ijk}$ is negative. These numbers are defined as
\begin{equation}
\alpha_{ijk} \equiv 2p_{i} + \sum_{l\neq j,k,i} p_l,\ (i\neq j, i\neq k, j\neq k).
\label{alpha}
\end{equation}
For the spacetimes with $d <  10$ one of the factors $\alpha$ is always negative and, hence one change
of Kasner regime is followed by another one, implying in such a way the oscillatory behaviour of the universe
in the neighbourhood of the cosmological singularity.
However, for the spacetimes with $d \geq 10$ exist such combinations of Kasner indices, satisfying  Eq. (\ref{gen-Kas1})
and   for which all the numbers $\alpha_{ijk}$ are positive. If a universe enters into the Kasner regime with such indices,
( so called ``Kasner stability region'') its chaotical behaviour disappers and this Kasner regime conserves itself.
Thus, the hypothesis was forward that in the spacetimes with $d \geq 10$ after a finite number of oscillations the universe
under consideration finds itself in the Kasner stability region and the oscillating regime is replaced by the monotonic Kasner behaviour.

The discovery of the fact that the chaotic character of the approach to the cosmological singularity disappears in the spacetimes with $d \geq 10$ was unexpected and looked as an accidental  result of a game between real numbers satisfying the generalized Kasner relations (\ref{new-Kas1}). Later it becomes clear that behind this fact there is a deep mathematical structure, namely, the hyperbolic Kac-Moody algebras. Indeed, in the series of works by Damour, Henneaux, Nicolai and some other authors (see e.g. Refs. \cite{DHN}) on the cosmological dynamics in the models based on superstring theories, living in 10-dimensional spacetime and on
the $d+1 = 11$ supergravity model, it was shown that in the vicinity of the singularity these models reveal ocsillating bahaviour of the BKL type. The important new feature of the dynamics in these models is the role played by non-gravitational bosonic fields
($p$-forms) which are also responsable for transitions from one Kasner regime to another. For description of these transitions
the Hamiltonian formalism \cite{Misner} becomes very convenient. In the framework of such formalism the configuration space of the
Kasner parameters describing the dynamics of the universe could be treated as a billiard while the curvature terms in
Einstein theory and also $p$-form's potentials in superstring theories play role of walls of these billiards. The transition from
one Kasner epoch to another is the reflection from one of the walls. Thus, there is a correspondence between rather complicated  dynamics of a universe in the vicinity of the cosmological singularity and the motion of an imaginary ball on the billiard table.

However, there exists a more striking and unexpected correspondence between the chaotical behaviour of the universe in the vicinity
of the singularity and such an abstract mathematical object as the hyperbolic Kac-Moody algebras \cite{DHN}. Let us explain briefly
what does it mean.
Every Lie algebra is defined by its generators ${h_i,e_i,f_i},i=1,\ldots,r$, where $r$ is the rank of the Lie algebra, i.e.
the maximal number of its generators $h_i$ which commutes each other (these generators constitute the Cartan subalgebra).
The commutation relations between generators are
\begin{eqnarray}
&&[e_i,f_j] = \delta_{ij}h_i,\nonumber \\
&&[h_i,e_j]  = A_{ij} = A_{ij}e_j,\nonumber \\
&&[h_i,f_j] = -A_{ij}f_j,\nonumber \\
&&[h_i,h_j] = 0.
\label{KM}
\end{eqnarray}
The coefficients $A_{ij}$ constitute the generalized Cartan $r \times r$  matrix such that
$A_{ii} = 2$, its  off-diagonal elements are non-positive integers and $A_{ij} = 0$ for $i\neq j$ implies
$A_{ji} = 0$. One can say that the $e_i$ are rising operators, similar to well-known operator $L_{+} = L_{x} + i L_{y}$
in the theory of angular momentum, while $f_i$ are lowering operators like $L_{-} = L_x - i L_{y}$. The generators $h_i$
of the Cartan subalgebra could be compared with the operator $L_z$. The generators should also obey the Serre's relations
\begin{eqnarray}
&&({\rm ad}\ e_i)^{1-A_{ij}} e_{j} = 0,\nonumber \\
&&({\rm ad}\ f_i)^{1-A_{ij}} f_j = 0,
\label{Serre}
\end{eqnarray}
where $({\rm ad} A) B \equiv [A,B]$.

The Lie algebras ${\cal G}(A)$ build on a symmetrizable Cartan matrix $A$ have been classified according to properties of their eigenvalues:\\
if $A$ is positive definite, ${\cal G}(A)$ is a finite-dimensional Lie algebra; \\
if $A$ admits one null eigenvalue and the others are all strictly positive ${\cal G}(A)$ is an Affine Kac-Moody algebra;\\
if $A$ admits one negative eigenvalue and all the others are strictly positive, ${\cal G}(A)$ is a Lorentz KM algebra.\\

There exists a correspondence between the structure of a Lie algebra and a certain system of vectors in the $r$-dimensional
Euclidean space, which simplify essentially the task of classification of the Lie algebras. These vectors called
roots represent the rising and lowering operators of the Lie algebra. The vectors corresponding to the  generators $e_i$ and
$f_i$ are called simple roots. The system of simple positive roots (i.e. the roots, corresponding to the rising generators
$e_i$) can be represented by nodes of their Dynkin diagrams, while the edges connecting (or non connecting) the nodes give
an informartion about the angles between simple positive root vectors.

An important subclass of Lorentz KM algebras can be defined as follows: A KM algebra such that the deletion of one node from its Dynkin diagram gives a sum of finite or affine algebras is called an hyperbolic KM algebra. These algebras are all known. In particular, there exists no hyperbolic algebra with rank higher than 10.

Let us recall some more definitions from the theory of Lie algebras. The reflections with respect to hyperplanes orthogonal to
simple  roots leave the systems of roots invariant. The corresponding finite-dimensional group is called Weyl group. Finally,
the hyperplanes mentioned above divide the $r$-dimensional Euclidean space into regions called Weyl chambers. The Weyl group
transform one Weyl chamber into another.

Now, we can briefly formulate the results of approach \cite{DHN} following the paper \cite{Damour}: the links between the billiards describing the evolution of the universe in the neigbourhood of singularity and its corresponding Kac-Moody algebra can be
described as follows:\\
the Kasner indices describing the ``free'' motion of the universe between the reflections from the wall correspond to the elements of the Cartan subalgebra of the KM algebra;\\
the dominant walls, i.e. the terms in the equations of motion responsible for the transition from one Kasner epoch to another, correspond to the simple roots of the KM algebra;\\
the group of refelections in the cosmological billiard is the Weyl group of the KM algebra;\\
the billiard table can be identified with the Weyl chamber of the KM algebra.\\

One can immagine two types of billiard tables: infinite such  where the linear motion without collisions with walls is possible
(non-chaotic regime) and those where reflections from walls are inevitable and the regime can be only chaotic. Remarkably, the Weyl
chambers of the hyperbolic KM algebras are designed in such a way that infinite repeating collisons with walls occur.
It was shown that all the theories with the oscillating approach to the singularity such as Einstein theory in dimensions $d < 10$
and superstring cosmological models correspond to hyperbolic KM algebras.

The existence of links between the BKL approach to the singularities and the structure of some infinite-dimensional Lie algebras has
inspired some authors to declare a new program of develpment of quantum gravity and cosmology \cite{DN}. They propose ``to take
seriously the idea that near the singularity (i.e. when the curvature gets larger than the Planck scale) the description of a spatial continuum and space-time based (quantum) field theory breaks down, and should be replaced by a much more abstract Lie algebraic description''.

\section{New types of cosmological singularities}
As was already mentioned in the Introduction, the development of the theoretical and observational cosmology and, in particular,
the discovery of the cosmic acceleration has stimulated the elaboration of  cosmological models where new types of singularities
were described. In contrast to the ``traditional'' Big Bang and Big Crunch singularities, these singularities occur not at zero
but at finite or even infinite values of the cosmological radius. The most famous of these singularities is, perhaps, the Big Rip singularity \cite{Rip,Star-Rip} arising if the absolute value of the negative pressure $p$ of the dark energy is higher than the energy density $\rho$.
Indeed, let us consider a flat Friedmann universe with metric
\begin{equation}
ds^2 - a^2(t)dl^2,
\label{Friedmann}
\end{equation}
filled with the perfect fluid with the equation of state
\begin{equation}
p = w\rho,\  w = const < -1.
\label{phantom}
\end{equation}
The dependence of the energy density $\rho$ on the cosmological radius $a$ is as usual
\begin{equation}
\rho = \frac{C}{a^{3(1+w)}}.
\label{phantom1}
\end{equation}
and
the Friedmann equation in this case has the form
\begin{equation}
\frac{\dot{a}^2}{a} = \frac{C}{a^{3(1+w)}},
\label{Friedmann1}
\end{equation}
where $C$ is a positive constant.
Integrating Eq. (\ref{Friedmann1}) we obtain
\begin{equation}
a(t) = \left(a_0^{\frac{3(1+w)}{2}} + \frac{2\sqrt{C}(t-t_0)}{3(1+w)}\right)^{\frac{2}{3(1+w)}}.
\label{Friedmann2}
\end{equation}
It is easy to see that at the finite moment $t_R > t_0$ equal to
\begin{equation}
t_R = t_0 - \frac{3(1+w)}{2\sqrt{C}} a_0 ^{\frac{3(1+w)}{2}}
\label{Rip}
\end{equation}
the cosmological radius becomes infinite and the same occurs also with the Hubble variable $\frac{\dot{a}}{a}$ and, hence,
with the scalar curvature. Thus, we encounter the new type of cosmological singularity, characterized by infinite values of
the cosmological radius, its time derivative, the Hubble variable and the scalar curvature. Usually this singularity is
called ``Big Rip'' singularity. Its properties have attracted an essential attention of researchers because of the fact that
some observational data indicate that the actual value of the equation of state parameter $w$ is indeed smaller than -1.

There are also other types of cosmological singularities which can be encountered at finite values of the cosmological radius
(see e.g. \cite{Brake,sudden,finite,boost,brane}). We shall consider here for illustration one type of singularities -
the Big Brake singularity \cite{Brake}. This singularity can be acchieved in a finite period of cosmic time and is characterized by
a finite value of the cosmological radius, by a vanishing first time derivative of the radius and by the second time derivative
of the cosmological radius tending to the minus infinity (an infinite deceleration). Let us consider a  perfect fluid
with the equation of state
\begin{equation}
p = \frac{A}{\rho} ,
\label{anti-Ch}
\end{equation}
where  $A$ is a positive constant. This fluid could be called ``anti-Chaplygin'' gas, because widely used the Chaplygin gas cosmological model \cite{Chap} is based on the equation of state $p = -A/\rho$.
The dependence of the energy density on the cosmological radius for the equation of state (\ref{anti-Ch}) is
\begin{equation}
\rho = \sqrt{\frac{B}{a^6}- A},
\label{anti-Ch1}
\end{equation}
where $B$ is a positive constant. When $a$ is small $\rho \sim 1/a^3$ and behaves like dust. Then , when $a \rightarrow a_B$
\begin{equation}
a_B = \left(\frac{B}{A}\right)^{\frac{1}{6}}
\label{anti-Ch2}
\end{equation}
and the energy density tends to zero. The solution of the Friedmann equation in this limit gives
\begin{equation}
a(t) = a_B - C_0(t_B-t)^{\frac43},\ C_0 = 2^{-7/3}3^{5/3}(AB)^{1/6}.
\label{anti-Ch3}
\end{equation}
Now, one can easily check that when $t \rightarrow t_B$ , $\dot{a} \rightarrow 0$ and $\ddot{a} \rightarrow -\infty$.
Thus, we indeed encounter the Big Brake cosmological singularity.

\section{Conclusions}
We have shown in this short review that the opinion expressed by L.D. Landau many years ago concerning
the importance of the problem of singularity in cosmology has proved to prophetic. The study of cosmological
singularity has revealed the existence of the oscillatory behaviour of a universe when the curvature of spacetime
grows, which in turn has a  deep connection with quite new branches of modern mathematics. From the other hand the last successes
of the observational cosmology has stimulated the development of various cosmological models, which reveal new types
of cosmological singularities, whose investigation from both physical and mathematical points of view can be very promising.

This work was partially supported by the RFBR grant 05-02-17450
and by the grant LSS-1157.2006.2.

\end{document}